\documentclass[pra,aps,amsmath,amssymb,showkeys]{revtex4}
\newcommand{\hh}{{\mathcal{H}}}

\newcommand{\lnp}{{\mathcal{L}}}
\newcommand{\lsp}{{\mathcal{L}}_{+}}
\newcommand{\lsa}{\mathcal{L}_{sa}}

\newcommand{\pen}{\openone}

\newcommand{\mm}{{\mathsf{M}}}
\newcommand{\tr}{\mathrm{tr}}
\newcommand{\id}{{\mathrm{id}}}

\newcommand{\dig}{{\mathrm{diag}}}
\newcommand{\bro}{\boldsymbol{\rho}}
\newcommand{\vbro}{\boldsymbol{\varrho}}

\newcommand{\zset}{\mathbb{C}}
\newcommand{\mset}{\mathbb{M}}
\newcommand{\rset}{\mathbb{R}}

\newcommand{\cla}{{\mathcal{A}}}
\newcommand{\clb}{{\mathcal{B}}}
\newcommand{\cle}{{\mathcal{E}}}
\newcommand{\clf}{{\mathcal{F}}}

\newcommand{\am}{{\mathsf{A}}}

\newcommand{\gx}{{\mathsf{G}}}
\newcommand{\me}{{\mathsf{E}}}
\newcommand{\sm}{{\mathsf{S}}}
\newcommand{\vm}{{\mathsf{V}}}
\newcommand{\um}{{\mathsf{U}}}
\newcommand{\wm}{{\mathsf{W}}}
\newcommand{\bmn}{{\mathsf{B}}}

\newcommand{\ax}{{\mathsf{X}}}
\newcommand{\ay}{{\mathsf{Y}}}
\newcommand{\vx}{{\mathsf{x}}}

\newcommand{\iu}{\mathtt{i}}

\begin{document}
\clearpage
\preprint{}

\title{On Kirkwood--Dirac quasiprobabilities and unravelings of quantum channel assigned to a tight frame}

\author{Alexey E. Rastegin}
\affiliation{Irkutsk State University, K. Marx St. 1,
Irkutsk 664003, Russia}

\begin{abstract}
An issue which has attracted increasing attention in contemporary
researches are Kirkwood--Dirac quasiprobabilities. List of their use
includes many questions of quantum physics. Applications of complex
tight frames in quantum information science were recently
demonstrated. It is shown in this paper that quasiprobabilities
naturally appear in the context of unravelings of a quantum channel.
Using vectors of the given tight frame to build principal Kraus
operators generates quasiprobabilities with interesting properties.
For an equiangular tight frame, we characterize the Hilbert--Schmidt
and spectral norms of the matrix consisted of quasiprobabilities.
Hence, novel uncertainty relations in terms of R\'{e}nyi and Tsallis
entropies are obtained. New inequalities for characterizing the
location of eigenvalues are derived. They give an alternative to
estimating on the base of Ger\v{s}gorin's theorem. A utility of the
presented inequalities is exemplified with symmetric informationally
complete measurement in dimension two.
\end{abstract}


\maketitle

\pagenumbering{arabic}
\setcounter{page}{1}

\section{Introduction}\label{ces1}

Equiangular tight frames are discrete sets of finite-dimensional
vectors with several notable properties \cite{lint66,seidel73}. Such
frames are interesting in their own right as well as for
applications in many disciplines such as signal and image
processing, data compression, sampling theory, and so on
\cite{caskut}. Complex equiangular tight frames with the maximal
number of elements lead to symmetric informationally complete
measurements. These measurements together with mutually unbiased
bases provide an indispensable tool in quantum information science.
Each complex tight frame provides the set of rank-one operators
forming a non-orthogonal resolution of the identity in the Hilbert
space. Measurements generated in this way seem to be very useful in
quantum information theory and deserve more attention than they have
obtained at the moment.

Dealing with an over-complete set of vectors results in many
distinctions from the most familiar case of quantum measurements in
orthonormal bases. New features can be characterized in terms of
quasiprobabilities assigned to one and the same measurement. In
general, quasiprobabilities represent quantum states similarly to
probability densities representing states in classical statistical
mechanics. Although the quasiprobabilities in a distribution sum to
one they can have negative and even non-real values. Wigner
functions \cite{wigner1932} are the well-known example of
quasiprobabilities used in various topics
\cite{casado08,vall11,vourdas2012,sels2013,elbaz2016,thlun}. The
Kirkwood--Dirac distribution is currently the subject of active
researches \cite{kirkwood33,dirac45}. It is closely related to the
so-called Terletsky--Margenau--Hill distribution
\cite{ypter37,hill61}. Recently, such distributions have found
applications in quantum state tomography
\cite{johan07,lundeen11,lundeen12,lundeen14}, information scrambling
\cite{halpern17,halpern18,halpern19,pollack19}, postselected
metrology \cite{dressel15,pusey19,lloyd20}, quantum thermodynamics
\cite{aea2014,miller17,levy20}, and studying conceptual questions
\cite{hofm15,halw16,drori21,bievre21}.

This paper explores some properties of tight frames expressed in
terms of the corresponding Kirkwood--Dirac quasiprobabilities
\cite{kirkwood33,dirac45}. These quasiprobabilities constitute a
Hermitian matrix assigned to a tight frame. Also, they can be
interpreted as related to Kraus operators of certain quantum
channel. The paper is organized as follows. Section \ref{ces2}
reviews the preliminary facts and gives the notation. Extremal
unravelings of quantum channels in general are examined in Section
\ref{ces3}. Kirkwood--Dirac quasiprobabilities assigned to a tight
frame are considered in Section \ref{ces4}. In Section \ref{ces5},
new entropic uncertainty relations are formulated for arbitrary
unraveling of the considered quantum channel. These results follow
from inequalities with the Hilbert--Schmidt and spectral norms of
the matrix consisted of the quasiprobabilities. The presented
relations are exemplified in Section \ref{ces6}. Section \ref{ces7}
concludes the paper with a summary of the results. Appendix
\ref{locva} is devoted to results concerning the location of
eigenvalues. An elementary inequality is proved in Appendix
\ref{appa}.

\section{Preliminaries}\label{ces2}

Let $\lnp(\hh)$ be the space of linear operators on
finite-dimensional Hilbert space $\hh$. By $\lsp(\hh)$ and
$\lsa(\hh)$, we respectively mean the set of positive semi-definite
operators and the real space of Hermitian ones. A state of the
quantum system of interest is represented by the density matrix
$\bro\in\lsp(\hh)$ normalized as $\tr(\bro)=1$. The set of pure
states contains density matrices of the form
$\bro=|\psi\rangle\langle\psi|$, where the ket $|\psi\rangle\in\hh$
is normalized as $\langle\psi|\psi\rangle=1$. For two operators
$\ax,\ay\in\lnp(\hh)$, their Hilbert--Schmidt inner product is
defined by the formula
\begin{equation}
\langle\ax,\ay\rangle_{\mathrm{hs}}=\tr(\ax^{\dagger}\ay)
\, . \label{hsip}
\end{equation}
Using some orthonormal basis as computational, vectors and operators
are represented by rectangular matrices. Let
$\mset_{m\times{n}}(\zset)$ be the space of all $m\times{n}$ complex
matrices. By $\mset_{n}(\zset)$, we mean the space of $n\times{n}$
complex matrices. The space of Hermitian $n\times{n}$ matrices is
denoted by $\mset_{n}^{(sa)}(\zset)$, and the set of positive
semi-definite $n\times{n}$ matrices is denoted by
$\mset_{n}^{+}(\zset)$. For each $\gx\in\mset_{m\times{n}}(\zset)$,
the square matrices $\gx^{\dagger}\gx$ and $\gx\gx^{\dagger}$ have
the same non-zero eigenvalues. The positive square roots of these
eigenvalues are the singular values $\sigma_{j}(\gx)$ of $\gx$
\cite{hj1990}. For real $q\geq1$, the Schatten $q$-norm is defined
as
\begin{equation}
\|\gx\|_{q}=\left(\sum\nolimits_{j}\sigma_{j}(\gx)^{q}\right)^{\!1/q}
\, , \label{qschat}
\end{equation}
where the sum is actually taken over non-zero singular values of
$\gx$. In particular, this family includes the trace norm for $q=1$,
the Hilbert--Schmidt norm, or the Frobenius norm,
\begin{equation}
\|\gx\|_{2}=\sqrt{\tr(\gx^{\dagger}\gx)}
 \label{qfro}
\end{equation}
for $q=2$, and the spectral norm
$\|\gx\|_{\infty}=\max\sigma_{j}(\gx)$. For
$\mm\in\mset_{n}^{+}(\zset)$, it holds that
\begin{equation}
\max\,\vx^{\dagger}\mm\,\vx=\|\mm\|_{\infty}
\, , \label{kyfn}
\end{equation}
where the maximum is taken over all normalized vectors
$\vx\in\mset_{n\times{1}}(\zset)$. The mentioned three norms
will often be used in the present paper.

Let us recall required material concerning tight frames in finite
dimensions \cite{caskut}. All the frames in $d$-dimensional Hilbert
space $\hh_{d}$ are assumed to be complex. A set
$\clf=\bigl\{|\phi_{j}\rangle\bigr\}_{j=1}^{n}$ of $n\geq{d}$ unit
kets of $\hh_{d}$ is called a frame if there exist strictly positive
numbers $S_{0}<S_{1}<\infty$ such that
\begin{equation}
S_{0}\leq\sum_{j=1}^{n}\bigl|\langle\phi_{j}|\psi\rangle\bigr|^{2}
\leq{S}_{1}
\label{abframe}
\end{equation}
for all unit $|\psi\rangle\in\hh_{d}$. The numbers $S_{0}$ and
$S_{1}$ are the minimal and maximal eigenvalues of the frame
operator
\begin{equation}
\sm=\sum_{j=1}^{n}|\phi_{j}\rangle\langle\phi_{j}|
\, . \label{sframe}
\end{equation}
We will further deal with the special case of tight frames, when
$S_{0}=S_{1}=S=nd^{-1}$ and $\sm=nd^{-1}\pen_{d}$. The states of a
tight frame induce the resolution
$\cle=\bigl\{\me_{j}\bigr\}_{j=1}^{n}$ of the identity so that
\begin{equation}
\frac{d}{n}\>\sm=\sum_{j=1}^{n} \me_{j}=\pen_{d}
\, . \label{sres}
\end{equation}
That is, the positive semi-definite operators
\begin{equation}
\me_{j}=\frac{d}{n}\>
|\phi_{j}\rangle\langle\phi_{j}|
 \label{sres1}
\end{equation}
form the so-called positive operator-valued measure (POVM). When the
pre-measurement state is described by density matrix $\bro$ with
$\tr(\bro)=1$, the probability of $j$-th outcome is equal to
\begin{equation}
\tr\bigl(\me_{j}\bro\bigr)=
\frac{d}{n}\,\langle\phi_{j}|\bro|\phi_{j}\rangle
\, . \label{prbj}
\end{equation}
Parseval tight frames obtained with $S=1$ are equivalent to
orthonormal bases commonly used in quantum theory.

Equiangular tight frames (ETFs) are especially interesting for many
reasons. The tight frame $\clf$ is called equiangular, when there
exists $c>0$ such that
\begin{equation}
\bigl|\langle\phi_{i}|\phi_{j}\rangle\bigr|^{2}=c
\label{md2c}
\end{equation}
for each pair $i\neq{j}$. Calculations then show that
\begin{equation}
c=\frac{S-1}{n-1}=\frac{n-d}{(n-1)d}
\ . \label{abeq}
\end{equation}
If there exists an ETF with $n$ elements in dimension $d$, then
$n\leq{d}^{2}$ and also exists an ETF with $n$ elements in
dimension $n-d$. The least case $n={d}^{2}$ with
\begin{equation}
\bigl|\langle\phi_{i}|\phi_{j}\rangle\bigr|^{2}=\frac{1}{d+1}
\qquad (i\neq{j})
 \nonumber
\end{equation}
gives a symmetric informationally complete measurement (SIC-POVM)
\cite{rbksc04}. Its existence for arbitrary $d$ is still an open
question, though exact constructions have been found
\cite{fhs2017,abb19,yard2020}.

\section{On extremal unravelings of quantum channels}\label{ces3}

The dynamics of systems in quantum information theory is typically
described in terms of Kraus operators \cite{nielsen}. Let us
consider a linear map
\begin{equation}
\bro\mapsto\Psi(\bro)=\sum\nolimits_{j}\am_{j}\bro \am_{j}^{\dagger}
\, , \label{brmap}
\end{equation}
where each of Kraus operators $\am_{j}$ maps kets of $\hh$ to kets
of $\hh^{\prime}$. In general, the input space $\hh$ and the output
space $\hh^{\prime}$ can differ. The map is called positive when it
maps elements of $\lsp(\hh)$ to elements of $\lsp(\hh^{\prime})$.
This property is clearly valid for (\ref{brmap}). In addition, maps
of the form (\ref{brmap}) are completely positive in the following
sense. Let us imagine an environmental system with its Hilbert space
$\hh^{\prime\prime}$. The complete positivity implies that the map
$\Psi\otimes\id^{\prime\prime}$ with identity map
$\id^{\prime\prime}$ is positive for arbitrary dimensionality of
$\hh^{\prime\prime}$. The considered map preserves the trace, when
its Kraus operators satisfy
\begin{equation}
\sum\nolimits_{j}\am_{j}^{\dagger}\am_{j}=\pen
\, , \label{brmap1}
\end{equation}
where $\pen$ is the identity operator on $\hh$. Trace-preserving
completely positive maps will be referred to as quantum channels.
They often called super-operators, where ``super'' conveys that the
map takes operators to operators \cite{preskill}. The concrete set
$\cla=\{\am_{j}\}$ in the right-hand side of (\ref{brmap}) will be
named an unraveling of the quantum channel. This terminology is due
to Carmichael \cite{carmichael} who introduced this word for a
representation of the master equation.

Probability distributions of interest will be characterized in terms
of the entropies of R\'{e}nyi \cite{renyi} and Tsallis
\cite{tsallis}. For $0<\alpha\neq1$, the R\'{e}nyi $\alpha$-entropy
and the Tsallis $\alpha$-entropy  are respectively defined as
\begin{align}
R_{\alpha}(\cla;\bro)&=\frac{1}{1-\alpha}\>
\ln\!\left(\sum\nolimits_{j} p_{j}(\cla;\bro)^{\alpha}
\right)
 , \label{renent}\\
H_{\alpha}(\cla;\bro)&=\frac{1}{1-\alpha}
\left(
\sum\nolimits_{j} p_{j}(\cla;\bro)^{\alpha}
- 1 \right)
=-\sum\nolimits_{j}p_{j}(\cla;\bro)^{\alpha}\,\ln_{\alpha}\bigl(p_{j}(\cla;\bro)\bigr)
\, , \label{tsaent}
\end{align}
where the probabilities are expressed as $p_{j}(\cla;\bro)=\tr\bigl(\am_{j}^{\dagger}\am_{j}\bro\bigr)$.
The so-called $\alpha$-logarithm of strictly positive $\xi$ is
defined as
\begin{equation}
\ln_{\alpha}(\xi)=
\begin{cases}
 \frac{\xi^{1-\alpha}-1}{1-\alpha}\>, & \text{ for } 0<\alpha\neq1 \, , \\
 \ln\xi\, , & \text{ for } \alpha=1 \,.
\end{cases}
\nonumber
\end{equation}
In the limit $\alpha\to1$, both the above entropies leads to the
Shannon entropy
\begin{equation}
H_{1}(\cla;\bro)=-\sum\nolimits_{j} p_{j}(\cla;\bro)\,\ln{p}_{j}(\cla;\bro)
\, . \label{shan1}
\end{equation}
It is easy to see that the Tsallis $\alpha$-entropy (\ref{tsaent})
is concave for all $\alpha>0$. Other properties of Tsallis
information functions and some physical applications are discussed,
e.g., in \cite{bplast,fyk04,sabe04,murty,majtey,blond,rastwork}. The
right-hand side of (\ref{renent}) is certainly concave for $\alpha\in(0,1)$
\cite{jizar2004}. Convexity properties of the R\'{e}nyi entropies
with orders $\alpha>1$ depend on dimensionality of probabilistic
vectors \cite{bengtsson,basrv1978}. For example, the binary
R\'{e}nyi entropy is concave for $0<\alpha\leq2$ \cite{basrv1978}.
However, this fact is not used in the following.

The complete sum of squared probabilities is usually referred to as
the index of coincidence
\begin{equation}
I(\cla;\bro)=\sum\nolimits_{j} p_{j}(\cla;\bro)^{2}
\, . \label{incon}
\end{equation}
In follows from (\ref{renent}) and (\ref{incon}) that
\begin{align}
R_{2}(\cla;\bro)&=-\,\ln{I}(\cla;\bro)
\, , \label{renent2}\\
H_{2}(\cla;\bro)&=1-I(\cla;\bro)
\, . \label{tsaent2}
\end{align}
The entropy (\ref{renent2}) is often referred to as the collision
entropy \cite{holik2013}. Another especially important case is the
min-entropy
\begin{equation}
R_{\infty}(\cla;\bro)=-\,\ln\bigl(\max{p}_{j}(\cla;\bro)\bigr)
\, , \label{reninf}
\end{equation}
which is obtained from (\ref{renent}) for $\alpha=\infty$.

It is well known that the choice of Kraus operators is not unique
due to unitary freedom in operator-sum representations. Two Kraus
representations of the same super-operator are related as
\begin{equation}
\bmn_{i}=\sum\nolimits_{j} \am_{j} u_{ji}
\, . \label{bxv}
\end{equation}
where the matrix $\um=[[u_{ji}]]$ is unitary \cite{nielsen,
preskill}. To make the two unravelings the same size, the smaller
set should be added by zero operators. Following \cite{rast2011},
one introduces the matrix $\Lambda(\cla;\bro)$ with entries
\begin{equation}
\bigl\langle\am_{i}\sqrt{\bro}\,,\am_{j}\sqrt{\bro}\,\bigr\rangle_{\mathrm{hs}}=
\tr\bigl(\am_{i}^{\dagger}\am_{j}\bro\bigr)
\, . \label{aijent}
\end{equation}
Hence, we immediately obtain the matrix relation
\begin{equation}
\um^{\dagger}\Lambda(\cla;\bro)\um=\Lambda(\clb;\bro)
\, . \label{vabg}
\end{equation}
The extremal unraveling $\cla^{(ex)}=\bigl\{\am_{i}^{(ex)}\bigr\}$
is obtained, when we take the unitary matrix $\vm$ that
diagonalizes $\Lambda(\cla;\bro)$, so that
\begin{align}
\vm^{\dagger}\Lambda(\cla;\bro)\vm&=\dig(\lambda_{1},\ldots,\lambda_{n})
\, , \label{vdeg}\\
\am_{i}^{(ex)}&=\sum\nolimits_{j} \am_{j} v_{ji}
\, . \label{vdeg1}
\end{align}
The eigenvalues listed in the right-hand side of (\ref{vdeg}) are
actually the probabilities $p_{i}(\cla^{(ex)};\bro)$. Hence, the
considered matrices are all positive semi-definite. Of course, the
unitary matrix $\vm$ depends on the actual density matrix of the
principal system. The obtained unraveling provides the extremality
property with respect to the Tsallis $\alpha$-entropies for
$\alpha\in(0,\infty)$ and the R\'{e}nyi $\alpha$-entropies for
$\alpha\in(0,1]$. Namely, for arbitrary unraveling $\clb$ we have
\begin{align}
 & H_{\alpha}(\clb;\bro)\geq{H}_{\alpha}(\cla^{(ex)};\bro)
& \forall{\ } \alpha\in(0;\infty)
\ , \label{tsaex} \\
 & R_{\alpha}(\clb;\bro)\geq R_{\alpha}(\cla^{(ex)};\bro)
& \forall{\ } \alpha\in(0;1]
\ . \label{renex}
\end{align}
Due to $\Lambda(\clb;\bro)=\wm\Lambda(\cla^{(ex)};\bro)\wm^{\dagger}$
with unitary $\wm$, one gets
\begin{equation}
p_{i}(\clb;\bro)=\sum\nolimits_{j}w_{ij}w_{ij}^{*}\,p_{j}(\cla^{(ex)};\bro)
\, . \label{pbpex}
\end{equation}
The results (\ref{tsaex}) and (\ref{renex}) follow from concavity of
the entropies and the fact that the square matrix with elements
$w_{ij}w_{ij}^{*}$ is unistochastic \cite{rast2011}. With respect to
the R\'{e}nyi $\alpha$-entropy, the unraveling with operators of the
form (\ref{vdeg1}) is generally extremal for $\alpha\in(0,1]$.
Nevertheless, the above unraveling allows us to estimate
$R_{\alpha}(\clb;\bro)$ from below for arbitrary unraveling $\clb$
and all $\alpha\geq2$. The corresponding new result is posed as
follows.

\newtheorem{tran0}{Proposition}
\begin{tran0}\label{pron0}
Let $\cla^{(ex)}$ be consisted of Kraus operators defined for the
given quantum channel and density matrix $\bro$ by (\ref{vdeg1}).
For arbitrary unraveling $\clb$ of this channel, it holds that
\begin{equation}
R_{\alpha}(\clb;\bro)\geq
\frac{\alpha-2}{\alpha-1}\,R_{\infty}(\cla^{(ex)};\bro)+\frac{1}{\alpha-1}\,R_{2}(\cla^{(ex)};\bro)
\, , \label{ain2}
\end{equation}
where $\alpha\in[2,\infty]$.
\end{tran0}

{\bf Proof.} Let us begin with the inequalities for $\alpha=2$ and
$\alpha=\infty$, namely
\begin{align}
R_{\infty}(\clb;\bro)&\geq R_{\infty}(\cla^{(ex)};\bro)
\, , \label{rinar}\\
R_{2}(\clb;\bro)&\geq R_{2}(\cla^{(ex)};\bro)
\, . \label{r2ar}
\end{align}
These relations are proved as follows. Due to (\ref{kyfn}), each of
the diagonal elements of positive semi-definite matrix $\mm$ is not
greater than $\|\mm\|_{\infty}$. Applying this to
$\Lambda(\clb;\bro)$ leads to the inequality
\begin{equation}
p_{i}(\clb;\bro)\leq\bigl\|\Lambda(\clb;\bro)\bigr\|_{\infty}=\bigl\|\Lambda(\cla^{(ex)};\bro)\bigr\|_{\infty}
\, . \label{pjam}
\end{equation}
Combining (\ref{reninf}) with (\ref{pjam}) completes the proof of
(\ref{rinar}).

The square of the Hilbert--Schmidt norm is equal to the sum of
squared moduli of all elements of the given square matrix. Hence, we
obtain
\begin{equation}
I(\clb;\bro)\leq\bigl\|\Lambda(\clb;\bro)\bigr\|_{2}^{2}=
\bigl\|\Lambda(\cla^{(ex)};\bro)\bigr\|_{2}^{2}=I(\cla^{(ex)};\bro)
\, . \label{r2ar2}
\end{equation}
Combining the latter with $R_{2}(\clb;\bro)=-\,\ln{I}(\clb;\bro)$
immediately gives (\ref{r2ar}).

Using (\ref{rinar}) and (\ref{r2ar}), we estimate
$R_{\alpha}(\clb;\bro)$ from below for intermediate values of
$\alpha$. It was proved in \cite{rastosid} that the R\'{e}nyi
$\alpha$-entropy of order $\alpha\in[2,\infty]$ satisfies
\begin{equation}
R_{\alpha}(\clb;\bro)\geq
\frac{\alpha-2}{\alpha-1}\,R_{\infty}(\clb;\bro)+\frac{1}{\alpha-1}\,R_{2}(\clb;\bro)
\, . \label{a2eq}
\end{equation}
The inequality (\ref{ain2}) directly follows from (\ref{rinar}),
(\ref{r2ar}) and (\ref{a2eq}). $\blacksquare$

Thus, the statement of Proposition \ref{pron0} allows one to
estimate $R_{\alpha}(\clb;\bro)$ from below in terms of
entropies of the unraveling with Kraus operators of the
form (\ref{vdeg1}). The novel relation (\ref{ain2}) has completed
in part the consideration of extremal unravelings given in
\cite{rast2011}. The analysis for $\alpha$ between $1$ and $2$ is an
open question. It seems that new methods should be developed to
resolve this question.

\section{Kirkwood--Dirac quasiprobabilities assigned to a tight frame}\label{ces4}

Let us begin with the concept of Kirkwood--Dirac quasiprobabilities
originally introduced for orthonormal bases. An extension to the
case of POVMs is posed as follows \cite{nqpo2021}. To the given POVM
$\cle=\bigl\{\me_{j}\bigr\}_{j=1}^{n}$ and density matrix $\bro$,
one assigns $n^{2}$ quantities of the form
$\tr\bigl(\me_{i}\me_{j}\bro\bigr)$. These quantities will be
referred to as generalized Kirkwood--Dirac quasiprobabilities
\cite{nqpo2021}. By $\Pi(\cle;\bro)$, we denote the $n\times{n}$
matrix constituted by these quasiprobabilities. In the case of POVM
with elements (\ref{sres1}), quasiprobabilities are expressed as
\begin{equation}
\tr\bigl(\me_{i}\me_{j}\bro\bigr)=
\frac{d^{2}}{n^{2}}\>
\langle\phi_{i}|\phi_{j}\rangle\langle\phi_{j}|\bro|\phi_{i}\rangle
\, . \label{qpkd}
\end{equation}
To each POVM, one can naturally assign trace-preserving completely
positive map
\begin{equation}
\bro\mapsto\Psi(\bro)=\sum\nolimits_{j=1}^{n}\am_{j}\bro\am_{j}^{\dagger}
\, , \label{mp1}
\end{equation}
where
\begin{equation}
\am_{j}=\sqrt{\me_{j}}
\, . \label{mp1a}
\end{equation}
The set $\cla=\{\am_{j}\}$ gives an unraveling of $\Psi$ in terms of
Kraus operators. The operators defined by (\ref{mp1a}) can be
treated as measurement operators in the sense of sections 2.2.3 and
2.2.6 of \cite{nielsen}. An ordinary link between measurement
operators and POVM elements is used here. Following \cite{brota19},
the operators $\am_{j}$ will be referred to as the principal Kraus
operators. Further, the above equations are rewritten as
\begin{align}
\bro\mapsto\Psi(\bro)&=\sum\nolimits_{j=1}^{n} p_{j}(\cla;\bro)\,|\phi_{j}\rangle\langle\phi_{j}|
\, , \label{map1}\\
\am_{j}&=\sqrt{\frac{d}{n}}\>|\phi_{j}\rangle\langle\phi_{j}|
\, , \label{map1a}\\
p_{j}(\cla;\bro)&=\frac{d}{n}\>\langle\phi_{j}|\bro|\phi_{j}\rangle
\, . \label{mp1b}
\end{align}
Quantum channels of the form (\ref{map1}) are particular examples of
entanglement breaking channels \cite{hshr}. Quantum channel is an
entanglement breaking one if and only if it has an unraveling with
rank-one Kraus operators \cite{hshr}.

In the following, we deal with the $n\times{n}$ matrix
$\Lambda(\cla;\bro)$ with elements
\begin{equation}
\tr\bigl(\am_{i}^{\dagger}\am_{j}\bro\bigr)=
\frac{d}{n}\>
\langle\phi_{i}|\phi_{j}\rangle\langle\phi_{j}|\bro|\phi_{i}\rangle
\, . \label{qpkd1}
\end{equation}
The latter differs from (\ref{qpkd}) only by a factor. Therefore, the
matrix equation
\begin{equation}
\Pi(\cle;\bro)=\frac{d}{n}\>\Lambda(\cla;\bro)
 \label{piam}
\end{equation}
is valid due to the chosen form of POVM elements. It is obvious that
$\Lambda(\cla;\bro)\in\mset_{n}^{(sa)}(\zset)$, whereas positive
semi-definiteness was mentioned right after (\ref{vdeg1}). If $n$
unit kets $|\phi_{j}\rangle$ generate a tight frame, then the form
(\ref{sres1}) of POVM elements implies (\ref{piam}). It is of
interest to ask, whether the implication holds in opposite
direction. The following statement takes place.

\newtheorem{tran1}[tran0]{Proposition}
\begin{tran1}\label{pron1}
Let elements of POVM $\cle=\bigl\{\me_{j}\bigr\}_{j=1}^{n}$ on
$\hh_{d}$ be all of rank one, and let quantum channel
$\Psi:\>\lnp(\hh_{d})\mapsto\lnp(\hh_{d})$ be assigned to $\cle$ in
accordance with (\ref{mp1}) and (\ref{mp1a}). The following two
statements are equivalent:
\begin{itemize}
  \item[(1)]{The POVM elements are represented as (\ref{sres1}), where $n$ unit kets $|\phi_{j}\rangle$ form a tight frame in $\hh_{d}$.}
  \item[(2)]{The matrix relation (\ref{piam}) holds for all unit vectors $|\psi\rangle\in\hh_{d}$.}
\end{itemize}
\end{tran1}

{\bf Proof.} The implication $(1)\Rightarrow(2)$ was actually shown
right before (\ref{piam}). One should prove that
$(2)\Rightarrow(1)$. Since the POVM $\cle$ consists of rank-one
elements $\me_{j}\in\lsp(\hh_{d})$ only, we can write
\begin{equation}
\me_{j}=\gamma_{j}^{2}\,|\phi_{j}\rangle\langle\phi_{j}|
\, , \label{mjph}
\end{equation}
where non-zero $\gamma_{j}\in\rset$ and each of $n$ kets
$|\phi_{j}\rangle$ is unit. Due to (\ref{mp1a}), the Kraus operators
read as $\am_{j}=\gamma_{j}\,|\phi_{j}\rangle\langle\phi_{j}|$. With
no loss of generality, we can take $\gamma_{j}>0$. Using
(\ref{mjph}) and $\bro=|\psi\rangle\langle\psi|$, the diagonal
elements of interest are represented as
\begin{equation}
\tr\bigl(\me_{j}^{2}\bro\bigr)=\gamma_{j}^{4}\,\bigl|\langle\phi_{j}|\psi\rangle\bigr|^{2}
\, , \qquad
\tr\bigl(\am_{j}^{2}\bro\bigr)=\gamma_{j}^{2}\,\bigl|\langle\phi_{j}|\psi\rangle\bigr|^{2}
\, . \label{twoen}
\end{equation}
Let the matrix relation (\ref{piam}) hold for all unit
$|\psi\rangle\in\hh_{d}$. Combining this with (\ref{twoen}) implies
that $\gamma_{j}^{2}=n^{-1}d$ for all $j=1,\ldots,n$. Further, the
completeness relation for $\cle$ then reads as
\begin{equation}
\frac{d}{n}\,\sum_{j=1}^{n} |\phi_{j}\rangle\langle\phi_{j}|=\pen_{d}
\, . \label{sres11}
\end{equation}
Hence, $n$ unit kets $|\phi_{j}\rangle$ form a tight frame in $\hh_{d}$.
$\blacksquare$

As was already mentioned, the choice of Kraus operators is not
unique due to unitary freedom in the operator-sum representation.
Using (\ref{bxv}) with the principal Kraus operators (\ref{map1a})
leads to another unraveling of the quantum channel. Kraus operators
of new unraveling are generally not of rank one. Being consisted of
the states of an ETF, the unraveling with the Kraus operators
(\ref{map1a}) plays a special role. It allows us to evaluate exactly
the Hilbert--Schmidt norms of the matrices $\Pi(\cle;\bro)$ and
$\Lambda(\cla;\bro)$. The following statement takes place.

\newtheorem{tran2}[tran0]{Proposition}
\begin{tran2}\label{pron2}
Let the matrix $\Lambda(\cla;\bro)$ with elements (\ref{aijent}) be
assigned to the given density matrix $\bro$ and ETF
$\clf=\bigl\{|\phi_{j}\rangle\bigl\}_{j=1}^{n}$; then
$\Lambda(\cla;\bro)\in\mset_{n}^{+}(\zset)$,
$\bigl\|\Lambda(\cla;\bro)\bigr\|_{1}=1$ and
\begin{equation}
\bigl\|\Lambda(\cla;\bro)\bigr\|_{2}^{2}=(1-c)\,I(\cla;\bro)+c\,\tr(\bro^{2})
\, . \label{tr2lam}
\end{equation}
\end{tran2}

{\bf Proof.} It was already mentioned that $\Lambda(\cla;\bro)$ is
positive semi-definite. Combining this fact with
$\tr\bigl(\Lambda(\cla;\bro)\bigr)=1$ implies
$\bigl\|\Lambda(\cla;\bro)\bigr\|_{1}=1$. To prove (\ref{tr2lam}),
we shall evaluate $\tr\bigl(\Lambda(\cla;\bro)^{2}\bigr)$ due to the
basic properties of equiangular tight frames. The $(i,i)$-entry of
the square of $\Lambda(\cla;\bro)$ reads as
\begin{align}
\frac{d^{2}}{n^{2}}\,\sum_{j=1}^{n}
\langle\phi_{i}|\phi_{j}\rangle\langle\phi_{j}|\bro|\phi_{i}\rangle
\langle\phi_{j}|\phi_{i}\rangle\langle\phi_{i}|\bro|\phi_{j}\rangle
&=\frac{d^{2}}{n^{2}}\,\langle\phi_{i}|\bro|\phi_{i}\rangle^{2}
+\frac{cd^{2}}{n^{2}}\,\sum_{j\neq{i}}\langle\phi_{i}|\bro|\phi_{j}\rangle\langle\phi_{j}|\bro|\phi_{i}\rangle
\label{stp1}\\
&=\frac{(1-c)d^{2}}{n^{2}}\,\langle\phi_{i}|\bro|\phi_{i}\rangle^{2}
+\frac{cd^{2}}{n^{2}}\,\sum_{j=1}^{n}\langle\phi_{i}|\bro|\phi_{j}\rangle\langle\phi_{j}|\bro|\phi_{i}\rangle
\nonumber\\
&=(1-c)\,p_{i}(\cla;\bro)^{2}+\frac{cd}{n}\,\langle\phi_{i}|\bro^{2}|\phi_{i}\rangle
\, . \label{stp2}
\end{align}
Here, the step (\ref{stp1}) follows from (\ref{md2c}) and the step
(\ref{stp2}) follows from (\ref{sres}). Summing (\ref{stp2}) over
all $i=1,\ldots,n$ finally gives
\begin{equation}
\bigl\|\Lambda(\cla;\bro)\bigr\|_{2}^{2}=\tr\bigl(\Lambda(\cla;\bro)^{2}\bigr)=(1-c)\,I(\cla;\bro)+c\,\tr(\bro^{2})
\, , \label{tr2lam1}
\end{equation}
where we again used (\ref{sres}) and (\ref{incon}).
$\blacksquare$

The Hilbert--Schmidt norm of $\Pi(\cle;\bro)$ is obviously expressed as
\begin{equation}
\bigl\|\Pi(\cle;\bro)\bigr\|_{2}=\sqrt{\frac{d}{n}}
\left((1-c)\sum\nolimits_{j=1}^{n}\tr\bigl(\me_{j}\bro\bigr)^{2}
+c\,\tr(\bro^{2})
\right)^{\!1/2}
 . \label{hstim}
\end{equation}
It should be pointed out that the result (\ref{tr2lam}) is proved
only for the concrete unraveling with the Kraus operators
(\ref{map1a}). Using a unitary freedom in the operator-sum
representation, we can extend a treatment to arbitrary unraveling of
the map (\ref{map1}). This will be done in the next section devoted
to uncertainty relations.

\section{Entropic uncertainty relations for unravelings assigned to an ETF}\label{ces5}

The above results lead to uncertainty relations for the quantum
channel defined by (\ref{map1}). Information entropies provide a
flexible tool to express uncertainties in quantum measurements. In
particular, this approach allows one to treat naturally uncertainty
relations in the presence of quantum memory
\cite{bccrr10,renes2009,ming2020,wang2022}. For a general discussion
of entropic uncertainty relations, see the review \cite{cbtw17} and
references therein. For some special type of measurements, entropic
uncertainty relations often follow from estimation of the
corresponding indices of coincidence. For measurements in mutually
unbiased bases, this approach was given in \cite{molmer,gedik2021}.
It is also useful for SIC-POVMs \cite{rastmubs} and their
generalizations \cite{rastsic}. Mutually unbiased bases have found
use in numerous questions \cite{durt}. In effect, they were used in
testing uncertainty relations for multiple measurements
\cite{ming2021}. It was mentioned in \cite{rastfram} that
equiangular tight frames deserve wider application in quantum
information science. The results of the present section aim to
support this claim.

Due to Proposition \ref{pron2}, we can evaluate the Hilbert--Schmidt
norm of all matrices of the form $\Lambda(\clb;\bro)$. These
positive semi-definite matrices have the same non-zero eigenvalues,
whence
\begin{equation}
\bigl\|\Lambda(\clb;\bro)\bigr\|_{q}=\bigl\|\Lambda(\cla;\bro)\bigr\|_{q}
\, . \label{ngqnq}
\end{equation}
It was proved in \cite{rastfram} that the index of coincidence
satisfies
\begin{equation}
I(\cla;\bro)\leq
\frac{Sc+(1-c)\,\tr(\bro^{2})}{S^{2}}
\ . \label{inbr}
\end{equation}
It can be shown that this inequality is saturated for density
matrices of the form
\begin{equation}
\vbro=\sum_{j=1}^{n}\nu_{j}\,|\phi_{j}\rangle\langle\phi_{j}|
\, , \label{nuj}
\end{equation}
where non-negative weights $\nu_{j}$ sum to $1$. In particular, the
equality takes place for the maximally mixed state
$\bro_{*}=d^{-1}\pen_{d}$ and for one of the pure states that form an
ETF $\clf$. For a SIC-POVM, the inequality (\ref{inbr}) is replaced
with equality for all density matrices. The corresponding result was
presented in \cite{rastmubs}.

Combining (\ref{tr2lam}) with (\ref{inbr}) immediately leads to the
inequality
\begin{equation}
\tr\bigl(\Lambda(\cla;\bro)^{2}\bigr)\leq\frac{(1-c)c}{S}+
\biggl(\frac{(1-c)^{2}}{S^{2}}+c\biggr)\tr(\bro^{2})
\, . \label{t2lam}
\end{equation}
It then follows from (\ref{t2lam}) and (\ref{dsc11}) that
\begin{align}
\left|\lambda_{j}-\frac{1}{n}\right|
&\leq\frac{\sqrt{n-1}}{n}\>\sqrt{n\,\tr\bigl(\Lambda(\cla;\bro)^{2}\bigr)-1}
\label{disc02}\\
&\leq\frac{\sqrt{n-1}}{n}
\>\sqrt{\biggl(\frac{(1-c)^{2}}{S^{2}}+c\biggr)n\,\tr(\bro^{2})+(1-c)cd-1}
\ . \label{disc2}
\end{align}
This interval is described in terms of purity $\tr(\bro^{2})$ and
the parameters of an ETF. In particular, we obtain
\begin{equation}
\bigl\|\Lambda(\cla;\bro)\bigr\|_{\infty}\leq\frac{1}{n}+\frac{\sqrt{n-1}}{n}
\>\sqrt{\biggl(\frac{(1-c)^{2}}{S^{2}}+c\biggr)n\,\tr(\bro^{2})+(1-c)cd-1}
\ . \label{ds2pr}
\end{equation}
Since matrices of the form $\Lambda(\clb;\bro)$ have the same
non-zero eigenvalues, the inequality (\ref{ds2pr}) holds for each of
these matrices. Hence, we have arrived at a conclusion.

\newtheorem{tran3}[tran0]{Proposition}
\begin{tran3}\label{pron3}
Let principal Kraus operators (\ref{map1a}) be built of the states
of ETF $\clf=\bigl\{|\phi_{j}\rangle\bigl\}_{j=1}^{n}$, and let
$\alpha\in[2,\infty]$. For arbitrary unraveling $\clb$ of the
channel (\ref{map1}) and each density matrix $\bro$, it holds that
\begin{align}
R_{\alpha}(\clb;\bro)&\geq
\frac{\alpha\ln{n}-2\ln{d}-\ln\bigl[(1-c)cS+\bigl((1-c)^{2}+cS^{2}\bigr)\tr(\bro^{2})\bigr]}{\alpha-1}
\nonumber\\
&-\frac{\alpha-2}{\alpha-1}\,\ln\!\left\{1+\sqrt{n-1}\,\sqrt{\bigl((1-c)^{2}S^{-2}+c\bigr)n\,\tr(\bro^{2})+(1-c)cd-1}\,\right\}
 . \label{ralp}
\end{align}
\end{tran3}

{\bf Proof.} It follows from (\ref{kyfn}) that
\begin{equation}
\underset{j}{\max}\,p_{j}(\cla^{(ex)};\bro)=\bigl\|\Lambda(\cla^{(ex)};\bro)\bigr\|_{\infty}
\, . \label{fpjb}
\end{equation}
Combining this with the definition of the min-entropy, (\ref{ngqnq}) and
(\ref{ds2pr}) immediately leads to the inequality
\begin{equation}
R_{\infty}(\cla^{(ex)};\bro)\geq\ln{n}-
\ln\!\left\{1+\sqrt{n-1}\,\biggl[\biggl(\frac{(1-c)^{2}}{S^{2}}+c\biggr)n\,\tr(\bro^{2})+(1-c)cd-1\biggr]^{1/2}\right\}
 . \label{ringa}
\end{equation}
By construction, the matrix $\Lambda(\cla^{(ex)};\bro)$ is diagonal. Due
to (\ref{ngqnq}) and (\ref{t2lam}), we then have
\begin{equation}
I(\cla^{(ex)};\bro)=\|\Lambda(\cla^{(ex)};\bro)\bigr\|_{2}^{2}=\|\Lambda(\cla;\bro)\bigr\|_{2}^{2}\leq
\frac{(1-c)c}{S}+
\biggl(\frac{(1-c)^{2}}{S^{2}}+c\biggr)\tr(\bro^{2})
\, , \label{t2lbm}
\end{equation}
whence
\begin{equation}
R_{2}(\cla^{(ex)};\bro)\geq2\ln{S}-
\ln\bigl[(1-c)cS+\bigl((1-c)^{2}+cS^{2}\bigr)\tr(\bro^{2})\bigr]
 . \label{ringb}
\end{equation}
Subsituting (\ref{ringa}) and (\ref{ringb}) into (\ref{ain2})
completes the proof. $\blacksquare$

The statement of Proposition \ref{pron3} gives a family of
uncertainty relations for an unraveling of quantum channel in terms
of R\'{e}nyi entropies. The inequality (\ref{t2lbm}) also leads to
Tsallis-entropy uncertainty relations. For $\alpha\in(0,2]$ and
arbitrary unraveling $\clb$, the Tsallis $\alpha$-entropy and the
index of coincidence satisfy
\begin{equation}
H_{\alpha}(\clb;\bro)\geq\ln_{\alpha}\bigl(I(\clb;\bro)^{-1}\bigr)
\, . \label{conc1}
\end{equation}
This inequality follows from Jensen's inequality and concavity of
the function $\xi\mapsto\ln_{\alpha}(\xi^{-1})$ for $0<\alpha\leq2$
\cite{rastmubs}. Combining (\ref{tsaex}) with (\ref{t2lbm}) and
(\ref{conc1}) leads to a conclusion.

\newtheorem{tran4}[tran0]{Proposition}
\begin{tran4}\label{pron4}
Let principal Kraus operators (\ref{map1a}) be built of the states
of ETF $\clf=\bigl\{|\phi_{j}\rangle\bigl\}_{j=1}^{n}$, and let
$\alpha\in(0,2]$. For arbitrary unraveling $\clb$ of the channel
(\ref{map1}) and each density matrix $\bro$, it holds that
\begin{equation}
H_{\alpha}(\clb;\bro)\geq
\ln_{\alpha}\!\left(\frac{S^{2}}{(1-c)cS+\bigl((1-c)^{2}+cS^{2}\bigr)\tr(\bro^{2})}\right)
 . \label{hralp}
\end{equation}
\end{tran4}

The statement of Proposition \ref{pron4} provides a family of
Tsallis-entropy uncertainty relations for an unraveling of the
quantum channel (\ref{map1}). Sometimes, Tsallis entropies are a
good alternative to R\'{e}nyi entropies. In particular, the case of
detection inefficiencies can be addressed in this way
\cite{rastmubs}. We refrain from presenting the details here.

Entropic uncertainty relations for a pair of unravelings were
studied in \cite{rast2011}. In contrast, the uncertainty relations
(\ref{ralp}) and (\ref{hralp}) are posed for a single unraveling.
Nevertheless, it is instructive to compare both the approaches
within a suitable example. To avoid a bulky presentation, we
restrict a consideration to state-independent uncertainty relations
for the case $n=d^{2}$. Another feature of uncertainty relations
derived in \cite{rast2011} is caused by the use of Riesz's theorem.
Namely, we have to deal with two different entropic parameters
constrained by a certain condition. To keep here a single entropy,
we will consider the Shannon one. It then follows from the results
of \cite{rast2011} that
\begin{equation}
H_{1}(\clb;\bro)\geq-\ln\bigl(\max\|\bmn_{i}\bmn_{j}\|_{\infty}\bigr)
\, . \label{h1aa}
\end{equation}
In general, the right-hand side of (\ref{h1aa}) is difficult to
evaluate. But this question is simplified for unraveling with
the principal Kraus operators (\ref{map1a}). In the case $n=d^{2}$
we obtain
\begin{equation}
H_{1}(\cla;\bro)\geq\ln{d}
\, . \label{ah1aa}
\end{equation}
The state-independent version of (\ref{hralp}) reads as
\begin{equation}
H_{\alpha}(\clb;\bro)\geq
\ln_{\alpha}\!\left(\frac{S^{2}}{(1-c)cS+(1-c)^{2}+cS^{2}}\right)
 . \label{hralpp}
\end{equation}
It holds for each unraveling $\clb$ of the channel (\ref{map1}) and
arbitrary quantum state. Calculating the right-hand side of
(\ref{hralp}) for $n=d^{2}$ leads to the inequality
\begin{equation}
H_{1}(\clb;\bro)\geq2\ln(d+1)-\ln(d+3)
\, . \label{hralb}
\end{equation}
The right-hand side of (\ref{ah1aa}) is slightly larger than the
right-hand side of (\ref{hralb}). But the former takes place for one
concrete unraveling, whereas the latter holds for all unravelings of
the channel (\ref{map1}). These uncertainty relations are
independent, but the latter is more important due to its scope. At
the same time, the difference between the right-hand sides of
(\ref{ah1aa}) and (\ref{hralb}) is enough small even for a space of
fewer dimensions, since
\begin{equation}
\ln(d^{2}+3d)-2\ln(d+1)=\ln\!\left(1+\frac{d-1}{(d+1)^{2}}\right)<\frac{d-1}{(d+1)^{2}}
\ . \nonumber
\end{equation}
With growth of $d$, the mentioned two bounds tend to coincide. This
example allows us to illustrate the advantage for utilizing the
current method.

\section{Examples of Kirkwood--Dirac quasiprobabilities}\label{ces6}

In this section, we briefly discuss explicit examples of the
matrices $\Lambda(\cla;\bro)$ and $\Pi(\cle;\bro)$ related via
(\ref{piam}). For an ETF-based measurement, the matrix
$\Lambda(\cla;\bro)$ consists of Kirkwood--Dirac quasiprobabilities
multiplied by $nd^{-1}$. For the maximally mixed state
$\bro_{*}=d^{-1}\pen_{d}$, we have
\begin{equation}
\Lambda(\cla;\bro_{*})=\frac{1}{n}
\begin{pmatrix}
    1 & c & \cdots & c \\
    c & 1 & \cdots & c \\
    \hdotsfor[2]{4} \\
    c & c & \cdots & 1
\end{pmatrix}
 . \label{latrix}
\end{equation}
Using $r_{i}=n^{-1}(n-1)c$ for all $i=1,\ldots,n$, we see from the
Ger\v{s}gorin theorem that
\begin{equation}
\left|\,\lambda_{j}-\frac{1}{n}\,\right|\leq\frac{(n-1)c}{n}=\frac{n-d}{nd}
\ . \label{ggdth}
\end{equation}
The inequality (\ref{inbr}) is saturated here with
$I(\cla;\bro_{*})=n^{-1}$. Combining this with $\tr(\bro_{*}^{2})=d^{-1}$
and (\ref{tr2lam}) leads to
\begin{equation}
\tr\bigl(\Lambda(\cla;\bro_{*})^{2}\bigr)=\frac{1-c}{n}+\frac{c}{d}
=\frac{d-1}{(n-1)d}+\frac{n-d}{(n-1)d^{2}}=\frac{d^{2}-2d+n}{(n-1)d^{2}}
\ , \label{tlam2}
\end{equation}
where we also used (\ref{abeq}). Meantime, direct calculations on
the base of (\ref{latrix}) give
\begin{equation}
\tr\bigl(\Lambda(\cla;\bro_{*})^{2}\bigr)=\frac{1+(n-1)c^{2}}{n}=
\frac{1}{n}\biggl(1+\frac{(n-d)^{2}}{(n-1)d^{2}}\biggr)=
\frac{nd^{2}-2nd+n^{2}}{n(n-1)d^{2}}
\ . \label{tlam22}
\end{equation}
In this way, we can checked our calculations because the quantities
(\ref{tlam2}) and (\ref{tlam22}) are equal. Substituting
$\tr(\bro_{*}^{2})=d^{-1}$ in the right-hand side of (\ref{disc2}),
we get (\ref{ggdth}) again. Therefore, for the maximally mixed state
our results lead to the same interval as Ger\v{s}gorin's theorem.

Taking one of pure states $|\phi_{j}\rangle$ leads to another case
that allows us to make calculations explicitly. For definiteness, we
further substitute $\bro=|\phi_{1}\rangle\langle\phi_{1}|$. The
matrix of interest then reads as
\begin{equation}
\Lambda(\cla;\bro)=\frac{d}{n}
\begin{pmatrix}
    1 & c & c &\cdots & c \\
    c & c & \ell_{23} & \cdots & \ell_{2n} \\
    c & \ell_{23}^{*} & c &\cdots & \ell_{3n} \\
    \hdotsfor[2]{5} \\
    c & \ell_{2n}^{*} & \ell_{3n}^{*} & \cdots & c
\end{pmatrix}
 . \label{patrix}
\end{equation}
Except for the first row and first column, off-diagonal complex
entries have moduli $|\ell_{ij}|=c^{3/2}$, whereas their phases
depend on the explicit structure of an ETF. The Ger\v{s}gorin
theorem establishes the two intervals, namely
\begin{align}
\left|\,\lambda_{j}-\frac{d}{n}\,\right|&\leq\frac{(n-1)cd}{n}
\ , \label{gers1}\\
\left|\,\lambda_{j}-\frac{cd}{n}\,\right|&\leq\frac{cd}{n}\,\bigl(1+(n-2)\sqrt{c}\,\bigr)
\, . \label{gers2}
\end{align}
Since the eigenvalues are all positive, these formulas can be
rewritten as
\begin{align}
\max\left\{2n^{-1}d-1,0\right\}\leq\lambda_{j}&\leq1
\, , \label{gerh1}\\
0\leq\lambda_{j}&\leq\frac{cd}{n}\,\bigl(2+(n-2)\sqrt{c}\,\bigr)
\, . \label{gerh2}
\end{align}
Using (\ref{gerh1}) and (\ref{gerh2}) to estimate the maximum of
positive eigenvalues, we have arrived at the inequality
\begin{equation}
\max\lambda_{j}=\bigl\|\Lambda(\cla;\bro)\bigr\|_{\infty}\leq1
\, , \label{specn1}
\end{equation}
with the constant right-hand side. The latter alone witnesses that
estimates could be improved. In contrast to (\ref{specn1}), the
inequality (\ref{disc2}) allows us to estimate $\max\lambda_{j}$
in terms of the parameters of an ETF. In the considered case, the
index of coincidence is equal to
\begin{equation}
I(\cla;\bro)=\frac{d^{2}}{n^{2}}+(n-1)\,\frac{c^{2}d^{2}}{n^{2}}
=\frac{d^{2}-2d+n}{n^{2}-n}
\ ,  \label{brpr}
\end{equation}
so that the inequality (\ref{inbr}) is saturated \cite{rastfram}.
Then the formulas (\ref{tr2lam}) and (\ref{disc02}) lead to
\begin{equation}
\left|\,\lambda_{j}-\frac{1}{n}\,\right|\leq\frac{1}{n}\>\sqrt{(1-c)(d^{2}-2d+n)+\frac{n^{2}}{d}-2n+1}
\ . \label{dsc2pur}
\end{equation}
As is shown in Appendix \ref{appa}, for $n>d$ the square root in the
right-hand side of (\ref{dsc2pur}) is less than $n-1$. Thus, one
has
\begin{equation}
0\leq\lambda_{j}<1
\, . \label{dsc2p}
\end{equation}
Hence, we deal with the inequality
\begin{equation}
\bigl\|\Lambda(\cla;\bro)\bigr\|_{\infty}\leq\frac{1}{n}+\frac{1}{n}\>\sqrt{(1-c)(d^{2}-2d+n)+\frac{n^{2}}{d}-2n+1}
\ , \label{dsc2pr}
\end{equation}
which is always better than the estimate on the base of (\ref{gerh1}).

To exemplify the obtained results, we consider symmetric
informationally complete quantum measurement in dimension two. Using
$\omega=\exp(\iu2\pi/3)$, one introduces the four kets
$|\phi_{j}\rangle$ such that
\begin{equation}
|\phi_{0}\rangle=|0\rangle
\, , \qquad
\sqrt{3}\,|\phi_{1}\rangle=|0\rangle+\sqrt{2}\,|1\rangle
\, , \qquad
\sqrt{3}\,|\phi_{2}\rangle=|0\rangle+\sqrt{2}\,\omega|1\rangle
\, , \qquad
\sqrt{3}\,|\phi_{3}\rangle=|0\rangle+\sqrt{2}\,\omega^{*}|1\rangle
\, . \nonumber
\end{equation}
Here, the vectors are numerated by indices from $0$, since this
better reflects the structure. The corresponding pure states are
represented on the Bloch sphere by vertices of tetrahedron. For
$\bro=|\phi_{0}\rangle\langle\phi_{0}|$, the matrix
(\ref{patrix}) reads as
\begin{equation}
\frac{1}{6}
\begin{pmatrix}
    3 & 1 & 1 & 1 \\
    1 & 1 & \frac{\iu}{\sqrt{3}}\, & -\frac{\iu}{\sqrt{3}} \\
        1 & -\frac{\iu}{\sqrt{3}} & 1 & \frac{\iu}{\sqrt{3}}\, \\
    1 & \frac{\iu}{\sqrt{3}}\, & -\frac{\iu}{\sqrt{3}} & 1
\end{pmatrix}
 . \label{satrix}
\end{equation}
Its non-zero eigenvalues are $2/3$ and $1/3$. Since $c=1/3$
according to (\ref{abeq}), we see from (\ref{dsc2pur}) that
\begin{equation}
\left|\,\lambda_{j}-\frac{1}{4}\,\right|\leq\frac{1}{4}\>\sqrt{\frac{11}{3}}
\ , \nonumber
\end{equation}
whence
\begin{equation}
0\leq\lambda_{j}<\frac{1}{4}+\frac{1}{4}\>\sqrt{\frac{11}{3}}<0.729
\, . \label{sic22}
\end{equation}
The Ger\v{s}gorin theorem gives here the two interval, whose union
leads to
\begin{equation}
0\leq\lambda_{j}\leq1
\, . \label{sic222}
\end{equation}
Both the ways are useless to bound non-negative eigenvalues from
below. But the result (\ref{sic22}) estimates the spectral norm of
(\ref{satrix}) from above much more precisely. The relative error is about
$9.3$ \% instead of $50$ \% due to (\ref{sic222}). For positive
semi-definite matrices of the considered form, the inequality
(\ref{dsc12}) deserves an attention together with the Ger\v{s}gorin
theorem.

\section{Conclusions}\label{ces7}

One considered Kirkwood--Dirac quasiprobabilities due to their
importance for quantum physics issues. Finite tight frames are
interesting in various fields including quantum information theory.
So, symmetric informationally complete measurements are a special
class of equiangular tight frames. This paper examined a quantum
channel with Kraus operators built of the vectors of an ETF. Then
Kirkwood--Dirac quasiprobabilities appear in the context of
unravelings of the quantum channel. Unitary freedom in the
operator-sum representation leads to a family of similar square
matrices. Each of them consists of the corresponding
quasiprobabilities. The revealed matrix properties allow one to
estimate from above some norms of the matrices of interest.

It is typical in statistical disciplines to examine links between
various quantitative characteristics. Say, entropic functions are
hardly exposed to measure immediately. The estimates derived in this
paper give a ground to examine desired connections. It was
exemplified by new uncertainty relations in terms of R\'{e}nyi and
Tsallis entropies for the considered quantum channel. We have also
obtained inequalities, which describe the location of eigenvalues of
positive semi-definite matrices. This provides an alternative to
estimation with the use of Ger\v{s}gorin's theorem. A symmetric
informationally complete measurement in dimension two illustrated a
significance of the obtained results. More detailed comparison of
the above alternatives would be a subject of separate investigation.

\appendix

\section{On the location of eigenvalues of a Hermitian matrix}\label{locva}

This appendix is devoted to the question how to characterize the
location of eigenvalues. The following statement takes place.

\newtheorem{mu1n}{Lemma}
\begin{mu1n}\label{lem1}
Let $\mm$ be a Hermitian $n\times{n}$ matrix, and let $\mu_{j}$ be one
of its eigenvalues. It holds that
\begin{equation}
\left|\,\mu_{j}-\frac{\tr(\mm)}{n}\,\right|
\leq\frac{\sqrt{n-1}}{n}\>\sqrt{n\|\mm\|_{2}^{2}-\tr(\mm)^{2}}
\, , \label{dsc11}
\end{equation}
with equality if and only if other eigenvalues of $\mm$ are all equal.
\end{mu1n}

{\bf Proof.} Recall that any Hermitian matrix is unitarily
diagonalizable and has real eigenvalues. For definiteness, we denote
the eigenvalues by $\mu_{j}$ and the eigenvalue of interest by
$\mu_{1}$. Let us put auxiliary values
\begin{equation}
x_{j}=\frac{\mu_{j}}{\tr(\mm)}
\label{xjdf}
\end{equation}
for all $j=1,\ldots,n$, so that these values sum to 1. As the
function $\xi\mapsto\xi^{2}$ is strictly convex, it follows from
Jensen's inequality that
\begin{equation}
\sum\nolimits_{j=1}^{n}x_{j}^{2}\geq{x}_{1}^{2}+\frac{(1-x_{1})^{2}}{n-1}=\frac{nx_{1}^{2}-2x_{1}+1}{n-1}
\ , \label{ix1n}
\end{equation}
with equality if and only if $x_{j}=(n-1)^{-1}(1-x_{1})$ for all
$j=2,\ldots,n$. The formula (\ref{ix1n}) can be rewritten as
\begin{equation}
(n-1)\left(n\sum\nolimits_{j=1}^{n}x_{j}^{2}-1\right)\geq{n}^{2}x_{1}^{2}-2nx_{1}+n-(n-1)
=(nx_{1}-1)^{2}
\, . \label{sqnx}
\end{equation}
Combining the latter with
\begin{equation}
\sum\nolimits_{j=1}^{n}x_{j}^{2}
=\frac{\tr(\mm^{2})}{\tr(\mm)^{2}}
\end{equation}
and $\tr\bigl(\mm^{2}\bigr)=\|\mm\|_{2}^{2}$ leads to the claim
(\ref{dsc11}) for $\mu_{1}$. Here, the condition for equality reads as
$\mu_{j}=(n-1)^{-1}\bigl(\tr(\mm)-\mu_{1}\bigr)$ for all
$j=2,\ldots,n$. $\blacksquare$

The condition of Hermiticity is essential for the above proof. For
arbitrary $n\times{n}$ matrix $\mm$ with complex entries, we have a
similar inequality in terms of singular values. Each singular value
$\sigma_{j}$ satisfies
\begin{equation}
\left|\,\sigma_{j}-\frac{\|\mm\|_{1}}{n}\,\right|
\leq\frac{\sqrt{n-1}}{n}\>\sqrt{n\|\mm\|_{2}^{2}-\|\mm\|_{1}^{2}}
\, , \label{dc11s}
\end{equation}
with equality if and only if other singular values of $\mm$ are all
equal. This statement can be shown in line with the proof of
(\ref{dsc11}). We refrain from presenting the details here.

The statement of Lemma \ref{lem1} allows us to characterize position
of eigenvalues of a Hermitian matrix on the real axis. It is an
alternative to more traditional way based on Ger\v{s}gorin's theorem
\cite{varga2004}. To a complex $n\times{n}$ matrix $\mm=[[m_{ij}]]$,
we assign $n$ positive numbers
\begin{equation}
r_{i}(\mm)=\sum_{j\neq{i}}|m_{ij}|
\, . \label{ridef}
\end{equation}
This term is the $i$-th deleted absolute row sum. For every
eigenvalue $\mu$ of $\mm=[[m_{ij}]]$ there is a positive integer $k$
such that \cite{varga2004}
\begin{equation}
\bigl|\mu-m_{kk}\bigr|\leq{r}_{k}(\mm)
\, . \label{gdth}
\end{equation}

Characterizing eigenvalues of positive semi-definite matrices is
often required in questions of quantum information theory. As a
rule, positive eigenvalues are then assumed to be arranged in
non-increasing order, viz.
\begin{equation}
\mu_{1}^{\downarrow}\geq\mu_{2}^{\downarrow}\geq\cdots\geq\mu_{n}^{\downarrow}\geq0
\, . \nonumber
\end{equation}
According to (\ref{dsc11}), the maximal eigenvalue of
$\mm\in\mset_{n}^{+}(\zset)$ is bounded from above as
\begin{equation}
\mu_{1}^{\downarrow}
\leq\frac{1}{n}\left(\|\mm\|_{1}+
\sqrt{n-1}\>\sqrt{n\|\mm\|_{2}^{2}-\|\mm\|_{1}^{2}}\,
\right)
 . \label{dsc12}
\end{equation}
This inequality is saturated if and only if other eigenvalues of
$\mm$ are all equal. The argumentation is the same as in the proof
of Lemma \ref{lem1}.

\section{An inequality}\label{appa}

This appendix aims to show that, for $n>d$, the square root in the
right-hand side of (\ref{dsc2pur}) is less than $n-1$. It will be
sufficient to prove
\begin{equation}
(1-c)(d^{2}-2d+n)+\frac{n^{2}}{d}-2n+1\leq
d^{2}-2d+\frac{n^{2}}{d}-n+1<(n-1)^{2}
\, , \label{aineq1}
\end{equation}
for all $n>d$. Indeed, one has $c\geq0$ with equality if and only if
$n=d$. The desired inequality can be rewritten as
\begin{equation}
f(n)=\frac{d-1}{d}\,n^{2}-n-d^{2}+2d>0
\, . \label{aineq2}
\end{equation}
Treating for a time $n$ as a continuous variable at fixed $d$, we
see a parabola $n\mapsto{f}(n)$ with $f(d)=0$. It intersects here
the abscissa axis with a positive slope due to
\begin{equation}
f^{\prime}(d)=2d-3>0
\, , \
\end{equation}
where integer $d\geq2$. Hence, the inequalities $f(n)>0$ and
(\ref{aineq1}) hold for all $n>d$.

\end{document}